\documentclass[runningheads]{llncs}
\usepackage{graphicx}
\usepackage{color}
\usepackage{amsmath}
\usepackage{booktabs}
\usepackage{amssymb}
\begin{document}
\title{ICS-Assist: Intelligent Customer Inquiry Resolution Recommendation in Online Customer Service for Large E-Commerce Businesses}
\titlerunning{ICS-Assist}
%
\author{Min Fu\inst{1,2} \and
Jiwei Guan\inst{2} \and
Xi Zheng\inst{2} \and
Jie Zhou\inst{1} \and
Jianchao Lu\inst{2} \and
Tianyi Zhang\inst{3} \and
Shoujie Zhuo\inst{1} \and
Lijun Zhan\inst{1} \and
Jian Yang\inst{2}}
\authorrunning{M. Fu et al.}
%
\institute{Alibaba Group, Hangzhou, China \and
Macquarie University, Sydney, Australia \and
Harvard University, Cambridge, Massachusetts, USA\\
\email{\{hanhao.fm,zj236040,souljoy.zsj,zhanlijun.zlj\}@alibaba-inc.com\\
\{james.zheng,jian.yang\}@mq.edu.au\\
\{jiwei.guan,jianchao.lu\}@hdr.mq.edu.au\\
\{tianyi\}@seas.harvard.edu
}
}
\maketitle 
\begin{abstract}
Efficient and appropriate online customer service is essential to large e-commerce businesses. Existing solution recommendation methods for online customer service are unable to determine the best solutions at runtime, leading to poor satisfaction of end customers. This paper proposes a novel intelligent framework, called ICS-Assist, to recommend suitable customer service solutions for service staff at runtime. Specifically, we develop a generalizable two-stage machine learning model to identify customer service scenarios and determine customer service solutions based on a scenario-solution mapping table. A novel knowledge distillation network called ``Panel-Student” is proposed to derive a small yet efficient distilled learning model. We implement ICS-Assist and evaluate it using an over 6-month field study with Alibaba Group. In our experiment, over 12,000 customer service staff use ICS-Assist to serve for over 230,000 cases per day on average. The experimental results show that ICS-Assist significantly outperforms the traditional manual method, and improves the solution acceptance rate, the solution coverage rate, the average service time, the customer satisfaction rate, and the business domain catering rate by up to 16\%, 25\%, 6\%, 14\% and 17\% respectively, compared to the state-of-the-art methods.

\end{abstract}
\section{Introduction}
Large e-commerce businesses such as Alibaba and Amazon provide hundreds of thousands of customer services to end customers via conversations every day, and these customer service conversations contain several topics, such as refunds, delivering inquiries, and instructions for using lucky money \cite{sari2018measuring}. When end customers make inquiries through online customer service, they usually demand their requirements and intentions be addressed as fast as possible \cite{KaufmanRon2015Wycs}. These requirements and intentions are usually versatile. As such, customer service solutions should be provided at runtime and should be able to correctly and timely address customers' requirements and intentions. For instance, when a customer calls in to complain about the poor quality of her newly bought shoes, we must recognize her intention of "returning the shoes" and provide her with the solution of how to return the shoes and apply for the refund \cite{KaufmanRon2015Wycs}.

Customer service solutions can be determined either manually or automatically. Determining customer service solutions manually is flexible and human-centric, and the representatives need to have enough expert knowledge to handle all types of customer problems \cite{mirchandani2012learning}. Several existing automated mechanisms have required expert knowledge learned from rich transaction history data to target most customer requirements. However, these approaches are inaccurate, inefficient and unsatisfactory, and most critically they are unable to generalize for diverse business domains \cite{rao2019bridging}. As such, end customers' satisfaction will be significantly affected, and business quality and profits will also be further influenced. 

In this paper, we propose a novel machine learning-based approach, called ICS-Assist, to facilitate customer service staff to identify ideal customer service solutions at runtime. ICS-Assist uses a two-stage learning model, coarse-grained learning and fine-grained learning, to identify the proper service scenario of each query made by the end customer. Moreover, ICS-Assist uses multi-aspect features (i.e. multi-round conversations, customer profiles, staff profiles, and order details) as the inputs to train a deep learning model for fine-grained service scenario recognition. Then ICS-Assist further determines the final solutions based on the ``scenario-solution" mapping table constructed by business operators. The main differences between our approach and existing methods are: 1) Our approach can achieve accurate customer service scenario recognition at runtime (i.e., while customer service staff are servicing end-customers); 2) We use a novel ``Panel-Student" learning scheme to derive a much smaller yet efficient learning model which can recognize service scenario at a finer granularity, a significant improvement over the traditional ``Teacher-Student" model \cite{hinton2015distilling}; 3) Our approach uses multi-aspect features instead of the commonly used language feature to train the ``Panel-Student" learning scheme and recognize service scenarios.

We implement ICS-Assist and evaluate it using a real-world field study with Alibaba Group. The experiments are conducted for over $6$ months. On average, over 12,000 customer service staff handle over 230,000 cases per day. We compare the performance of ICS-Assist with existing semantic and relevance matching methods, including HCAN \cite{rao2019bridging}, ESIM-seq \cite{chen2017enhanced}, DAM \cite{zhou-etal-2018-multi}, and DIIN \cite{DBLP:conf/iclr/GongLZ18}. The experimental results are two-fold: 1) Our method increases the solution acceptance rate by up to 16\%, increases solution coverage rate by up to 25\%, reduces average service time by up to 6\%, increases customer satisfaction rate by up to 14\%, and increases business domain catering rate by up to 17\%, compared to the state-of-the-art methods; 2) Our method increases the solution acceptance rate by 24\%, increases solution coverage rate by 34\%, reduces average service time by 8\%, increases customer satisfaction rate by 19\%, and increases business domain catering rate by 22\%, compared to the traditional manual method. 


The research contributions of this paper are 1) We propose a novel intelligent framework to recognize customer service scenarios and further determine appropriate customer service solutions
at runtime. In this way, we extend the idea of the ``Teacher-Student" model to propose a generalizable ``Panel-Student" distilled learning method that determines suitable customer service scenarios and solutions for multiple e-commerce business domains. 2) We show a real-world field study to demonstrate the efficacy and validity of our proposed approach.

The remainder of this paper is as follows: Section \ref{sec:background} introduces the background; Section \ref{sec:method} illustrates our proposed approach; Section \ref{sec:experiment} describes the experimental evaluation; Section \ref{sec:threat} discusses threats to validity; Section \ref{sec:relatedwork} provides related work; Section \ref{sec:conclusion} provides the conclusion and future work.

\section{Background}
\label{sec:background}
\subsection{Intelligent Customer Service in E-Commerce }
 E-commerce customer service plays a significant role in business 
 profit-making
 and customer satisfaction \cite{sari2018measuring}. 
 In contrast to traditional customers' service involving huge human efforts,
 organizations use intelligent customer service 
 to promote
 effortless customers experiences 
 and 
 improve productivity.
 Specifically, the state-of-the-art intelligent customer service is not just multi-channel but omnichannel, which allows the organizations to facilitate effective interactions between them and their customers by unifying the experience across self-assisted and field-service channels \cite{Intelligent_Customer_Service}. In large e-commerce corporations, such as Alibaba, JD.Com and Amazon, intelligence customer service has been successfully used to save their customer service costs by over 20\%. With these successful stories, many small to medium-sized e-commerce companies are starting to develop their intelligent customer service systems \cite{KaufmanRon2015Wycs}.

\subsection{Business Requirements for Customer Service}
\label{subsec:business_req_for_cs}
As a critical component of the business chain, customer service has been regularized by standardized business requirements, which are formulated by several popular e-commerce corporations based on over 20 years' business exploration \cite{eales2012impact}. 
These requirements are 1) Customer service solutions should be correctly determined; 2) The customer service system should cover as many customer service solutions as possible; 3) The time spent on customer service dialogues should be minimized; 4) The satisfaction rate of end customers should be maximized; 5) The Customer service system should be able to cater for as many business domains as possible. Hence, the e-commerce industry uses the following business 
metrics to evaluate the quality of customer service: 1) Solution Acceptance Rate (SAR), which refers to the percentage of solutions that are accepted by end customers; 2) Solution Coverage Rate (SCR), which refers to the proportion of the solutions that can be recalled from the overall solutions; 3) Average Service Time (AST), which refers to the average time spent on customer service conversations; 4) Customer Satisfaction Rate (CSR), which refers to the percentage of the customers who are satisfied with the customer service; 5) Business Domain Catering Rate (BCR), which refers to how many business domains can be catered for by the customer service system.

\section{Our Proposed Method}
\label{sec:method}
\begin{figure}[!h]
\vspace{-3mm}
\centering
\includegraphics[width=0.91\columnwidth]{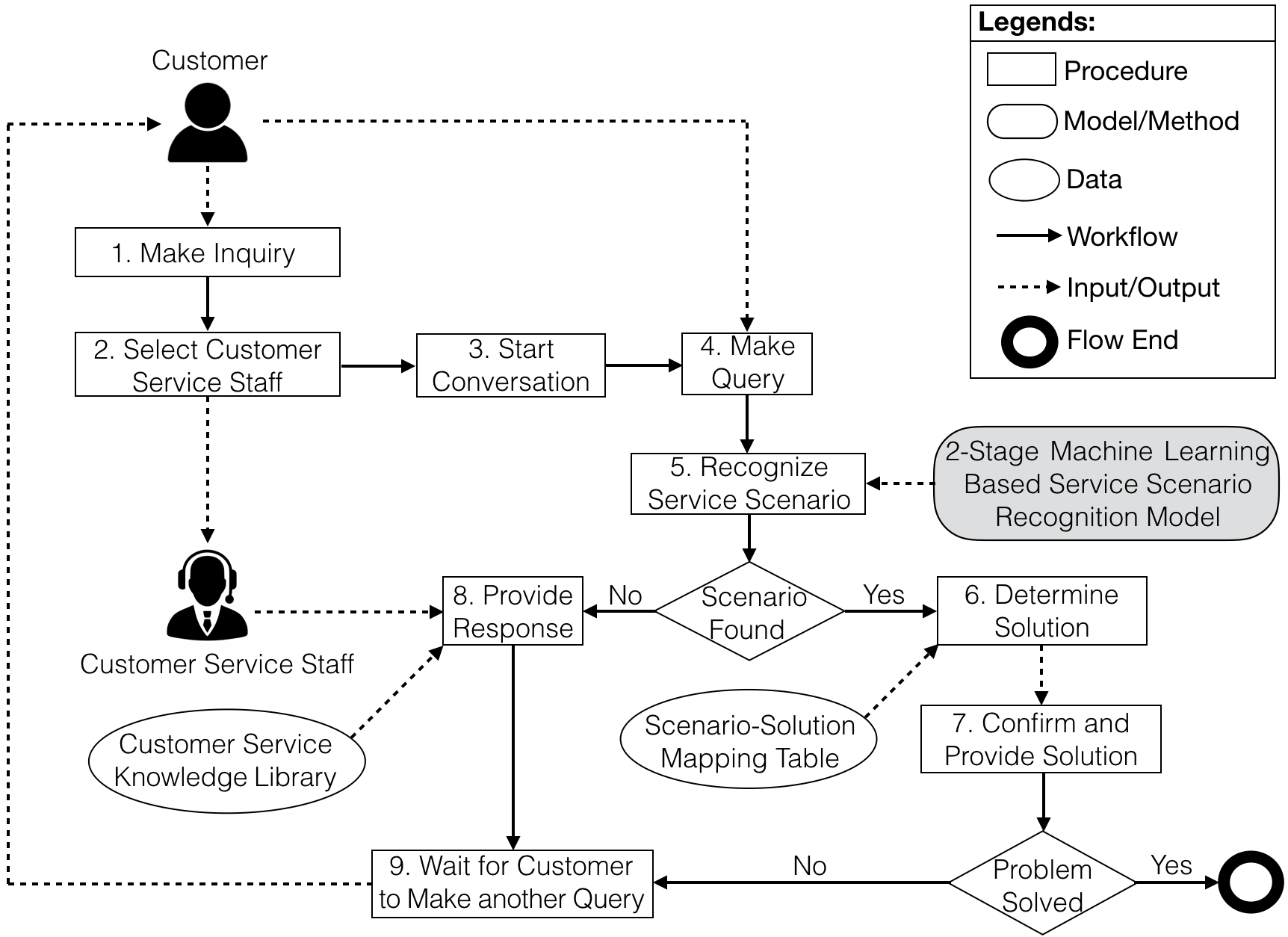}
\caption{Overview of ICS-Assist}
\vspace{-3mm}
\label{fig:figure_2}
\end{figure}

Our approach is based on the following design decisions: 1) Service solutions should be mapped from recognized service scenarios based on the well-established ``scenario to solution" mapping rules defined by the e-commerce business; 2) Customer service scenarios must be determined in a runtime manner; 3) The model can utilize a multi-stage paradigm in order to recognize the optimal customer service scenarios to determine the optimal service solutions. The overview of our proposed approach, named ICS-Assist, is shown in Fig. \ref{fig:figure_2}. When a customer inquires, the system selects an available customer service staff to start the conversation. After the customer makes each query, ICS-Assist recognizes the relevant service scenarios based on the two-stage machine learning (coarse-grained learning and fine-grained learning) scenario recognition model proposed by us. If scenarios are not found, the customer service staff responds to the customer on her own expert experience; otherwise, ICS-Assist determines the solutions based on the scenario-solution mapping table maintained by the business itself, and the customer service staff confirms and provides the solutions to the customer. If the problem of the end customer is solved, the customer service ends; otherwise, ICS-Assist waits for the end customer to make another query, and the aforementioned procedure repeats until the problem is solved.

\subsection{Data Preparation \& Preprocessing}

\begin{figure}[!h]
\centering
\includegraphics[width=0.85\columnwidth]{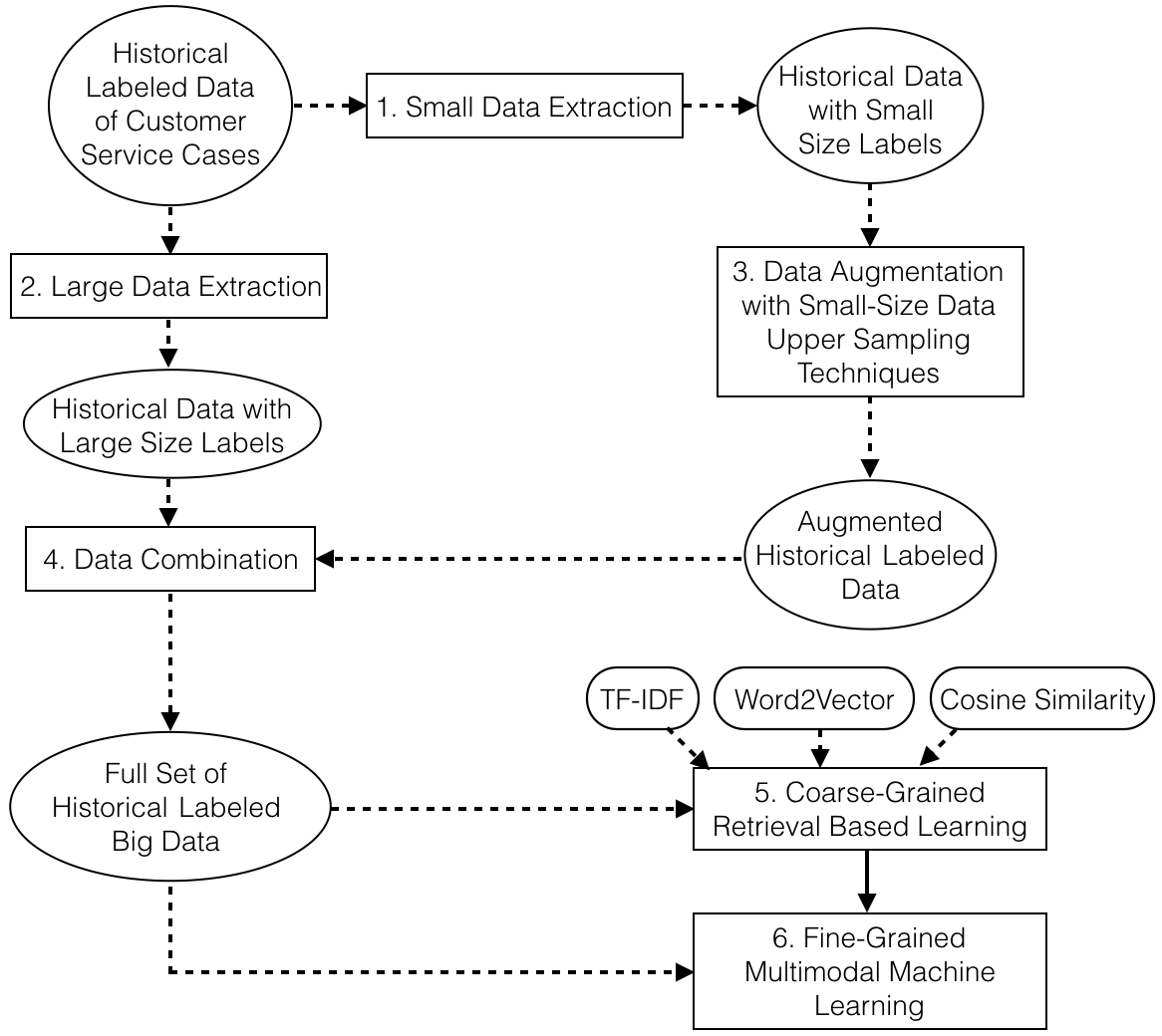}
\caption{Training Process for Service Scenario Recognition Model}
\label{fig:figure_3}
\end{figure}

The data processing pipeline for the service scenario recognition in ICS-Assist is shown in Fig. \ref{fig:figure_3}. The input data is generated from the historical customer service log, which contains  customer utterances and staff operations (e.g. clicking, hovering and querying) in a service session. The service scenarios clicked or searched by the staff are paired with the customer utterances to form the positive samples in the dataset. We also manually check these automatically generated pairs.

However, the training set is imbalanced as some regular service scenarios have millions of cases, such as inquiries about a delivery, refunding, while others may only contain a few thousand. Thus the corpus of customer utterance can be too sparse to learn a well-generalized model. To address this, we apply the data augmentation method of up-sampling on the scarce cases to expand their original size to $100$ times. 
Finally, we combine the augmented rare cases with the regular cases as the positive samples, and also randomly choose an equal number of irrelevant pairs of service scenarios and customer utterances as negative samples.

The data structure for each training sample is a triplet consisting of the customer utterance $\mathcal{U}$, the description of standard service scenario $\mathcal{S}$ and label $y$, where  $y\in\{0, 1\}$, both $\mathcal{U}$ and $\mathcal{S}$ are text, $\mathcal{U} = (w^u_1, w^u_2, w^u_3, ...), \mathcal{S} = (w^s_1, w^s_2, w^s_3, ...)$, $w_i$ is the $i$-th word in the sequence.

The model learning follows a two-stage procedure that contains the coarse-grained ranking and the fine-grained ranking. At the coarse-grained ranking stage, we use a simple approach that narrows down the search range in the candidate set to filter out the irrelevant scenarios. At the fine-grained ranking stage, we propose a Panel-Student knowledge distillation approach to train a lighter model that is able to find out the most suitable service scenarios.

\subsection{Coarse-Grained Learning Model}
We describe the process for the coarse-grained model. First, we compute the representation $\mathbf{u}$, $\mathbf{s}$ for $\mathcal{U}$, $\mathcal{S}$:

\begin{align}
    \sum_i \mathrm{tf\_idf}(w_i) \times \mathrm{word2vec}(w_i)
\end{align}

where $u$, $s$ $\in$ $\mathbb{R}^{d\_\{wv\}}$, d\_\{wv\} is the dimensionality for Word2Vec \cite{mikolov2013distributed}. 
The representation is exactly the weighted average of the word vector for the corresponding words in the text, where the weight we use here is tf-idf.

We get the top-$K$ suitable scenarios by comparing cosine similarity $cos\_sim(\mathbf{u}, \mathbf{s}_k)$, where $\mathbf{s}_k$ is the representation for a scenario in the candidate set. After that, the top-$K$ candidates would be fed into the fine-grained model.

\subsection{Fine-Grained Learning Model}
The fine-grained model learns complicated semantic relationships between customer utterances and service scenario descriptions and finds out the most suitable service scenarios. In our case, the ranking model requires very high precision. To achieve this goal, the simplest way is that we train a model as large as possible with strong fitting capacities, but by doing so the inference time would be slower, which is unfriendly for online recommendation at runtime.

Knowledge distillation \cite{hinton2015distilling,you2017learning} is an effective way to distill the knowledge learned from the teacher model and builds an accurate lightweight student model. The teacher model is usually a large neural network or an ensemble of networks containing millions parameters. Hence, the state-of-the-art large-scale pre-trained language models, such as ELMO \cite{peters-etal-2018-deep}, BERT \cite{devlin-etal-2019-bert} and XLNet \cite{yang2019xlnet}, can serve as the teacher network. These models are millstones in natural language processing filed and significantly improve the performance of many downstream tasks such as question answering, textual entailment, and text classification etc.




However, among the aforementioned pre-trained language models, using only one of them as a teacher network seems to be unable to completely train a generalizable student model that can achieve as good performance as the teacher network in diverse business domains. Besides, our empirical studies show that ELMO has the best performance and is slightly better than BERT in the business domain of Alibaba Movie (which is an Alibaba business portal for watching online movies), while in the business domain of Tmall Global (is an Alibaba web portal for selling imported commodities), ELMO's performance is the worst among the three pre-trained language models. Thus, in order to cater for all types of  business domains, we explore a Panel-Student knowledge distillation approach that combines all the three {\em teachers} to form a generalizable {\em panel}.

\begin{figure}[!h]
\vspace{-5mm}
\includegraphics[width=\columnwidth]{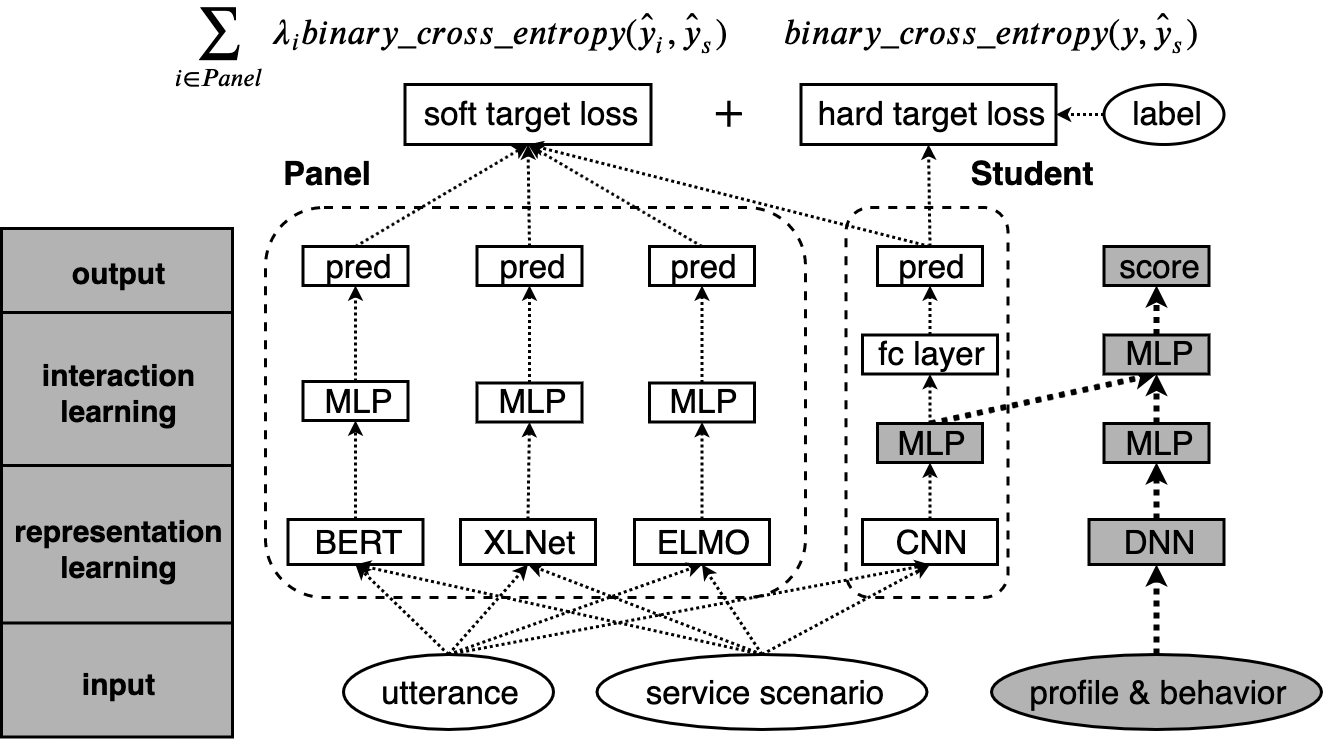}
\caption{Fine-Grained ``Panel-Student" Learning Model}
\label{fig:figure_4}
\vspace{-10mm}
\end{figure}

The full details of the fine-tuning ``Panel-Student" learning scheme are illustrated in Fig. \ref{fig:figure_4}, which can be divided into four layers:

\begin{enumerate}
    \item the input layer that maps the raw text data into the word embeddings;
    \item the representation learning layer 
    that encodes the word embeddings into a comprehensive contextualized representation;
    \item the interaction learning layer 
    that further processes the representation and extracts both semantic-oriented and relevance-oriented matching signals between the input utterance and service scenario; 
    \item the output layer that generates the final matching scores.
\end{enumerate}

\subsubsection{Panel-Student Framework}
We use three high-capacity pre-trained language models in our panel, including ELMO \cite{peters-etal-2018-deep}, BERT-LARGE \cite{devlin-etal-2019-bert}, and XLNet \cite{yang2019xlnet}. As the core task of our ICS-Assist is to match the customer utterance with a suitable service scenario by leveraging semantic-oriented and relevance-oriented matching signals in the text pairs, we follow the fine-tuning setting of the text-entailment task described in each corresponding teacher model (text-entailment can be viewed as a special case of text match \cite{rao2019bridging}). After the fine-tuning stage, we train the student model under the panel's supervision with the soft target loss (shown in Figure \ref{fig:figure_4}) and the hard target loss with ground truth labels. We choose TextCNN \cite{sun2018new} as the student model due to its lightweight and fast inference.

The input layer maps the words within $\mathcal{U}$ and $\mathcal{S}$ into the corresponding embeddings $U = [\mathbf{w}^u_1; \mathbf{w}^u_2; \mathbf{w}^u_2; ... \mathbf{w}^u_N]$ and 
$S = [\mathbf{w}^s_1; \mathbf{w}^s_2; \mathbf{w}^s_2; ... \mathbf{w}^s_N]$, where $\mathbf{w}_i \in \mathbb{R}^{d}$ is the corresponding embedding given a word $w_i$, $U, S \in \mathbb{R}^{N \times d}$. We pad the variable length sequence to fixed-length $N$.

At the representation learning stage, 
we first apply 1-d convolution over $U$ with $k$ different kernel size:
\begin{equation}
\bar{U}^k = \sigma(W_f^k \ast U + b^k),
\end{equation} where $W^k_f$ is the parameter for $k$-th convolution kernel, $b^k$ is the bias, $\sigma$ is the activation function, the output channel number for convolution is $d_o$, $\bar{U}^k \in \mathbb{R}^{N \times d_o}$. Then we take the maximum and average values over the sequence length dimension and concatenate them to form an overall representative semantic signal. Taking the maximum value can effectively extract the features for the occurrence of some keywords and taking the average can be more robust to the noise in the corpus.
\begin{align}
\mathbf{\bar{u}}^k_{max} & = \mathrm{max}(\bar{U}^k), \\ \mathbf{\bar{u}}^k_{mean} & = \mathrm{mean}(\bar{U}^k), \\
\mathbf{\tilde{u}} & = [
\mathbf{\bar{u}}^1_{max}; 
...;
\mathbf{\bar{u}}^k_{max};
\mathbf{\bar{u}}^1_{mean};
...;
\mathbf{\bar{u}}^k_{mean}],
\end{align} 
where $\mathbf{\bar{u}}^k_{max}, \mathbf{\bar{u}}^k_{mean} \in \mathbb{R}^{d_o}$, $ \mathbf{\tilde{u}} \in \mathbb{R}^{2kd_o} $. The representation $\mathbf{\tilde{s}}$ for $\mathcal{S}$ is obtained in a similar way.

At the interaction learning stage, we enhance the interaction between $\mathcal{U}$ and $\mathcal{S}$ by applying more complicated arithmetic operations on the original signal obtained in the previous stage. The original signals $\mathbf{\tilde{u}}$,  $\mathbf{\tilde{s}}$
, the element-wise multiplication of the original signals, the element-wise square of the difference between 
the two signals
are concatenated together and fed to an MLP to generate the final matching feature $\mathbf{m}$:

\begin{align}
    \mathbf{x} & = [\mathbf{\tilde{u}}; \mathbf{\tilde{s}}; \mathbf{\tilde{u}} \otimes \mathbf{\tilde{s}} ; (\mathbf{\tilde{u}} - \mathbf{\tilde{s}}) ^ {\circ 2}], \\
    \mathbf{m} & = \mathrm{MLP}(\mathbf{x}),
    \label{eq:mlp}
\end{align}

where $\mathbf{x} \in \mathbb{R}^{8kd_o}$.
After the student model is trained, the model can make the following prediction:
\begin{align}
    g^s = \sigma(W\mathbf{m} + b), \\
    \hat{y}_s = \mathrm{sigmoid}(g^s),    
\end{align}
where $\sigma$ is the activation function, and $\hat{y}_s$ is a real number between 0 and 1, which represents the probability of standard service scenario $\mathcal{S}$ matching the given utterance $\mathcal{U}$.\\

\subsubsection{Final Hybrid Model for Service Scenario Recognition}

As shown at the rightmost side in Figure \ref{fig:figure_4}, we also use multi-aspect features $\mathbf{\bar{m}}$ learned by a DNN model based on the customer profiles, the log of historical customer behavior and customer service staff. The intermediate multi-aspect feature $\mathbf{\bar{m}}$ will be combined with $\mathbf{m}$ (see equation \ref{eq:mlp}) generated by the student TextCNN, and then fed to an MLP to make the final prediction as follows:

\begin{align}
    g^h & = \mathrm{MLP}([\mathbf{m}; \mathbf{\bar{m}}]), \\
    \hat{y}_h & = \mathrm{sigmoid}(g^h).
\end{align}

The training of the final hybrid model which utilizes the output from our student model has three phases. The first two phases are used for training the student model alone, and the third phase is used for training the hybrid model:
\begin{enumerate}
    \item 
    Each teacher model in the panel is fine-tuned;
    \item 
    The student model (i.e., TextCNN) is trained within the ``Panel-Student" scheme using the loss function below:
    \begin{equation}
    \sum_{i \in Panel}\lambda_i \mathrm{binary\_cross\_entropy}(\hat{y}_i, \hat{y}_s) + \mathrm{binary\_cross\_entropy}(y, \hat{y}_s),
    \end{equation}

    The lost function consists of two types of loss: 1) the soft-target loss between the student model's predictions $\hat{y_s}$ and each teacher model's predictions $\hat{y_i}$; 2) the hard target loss between the predictions of the student model $\hat{y_s}$ and the ground truth labels $y$.  

    \item 
  All layers within TextCNN up to the MLP are extracted and combined with the output from the MLP layer within the DNN model at the rightmost in Figure \ref{fig:figure_4} to construct the final hybrid model for scenario prediction. The hybrid model is trained using this loss function:
    \begin{align}
        \mathrm{binary\_cross\_entropy}(y, \hat{y}_h),
    \end{align}
    where $\hat{y}_h$ is the prediction for the hybrid model.
\end{enumerate}

\subsection{Scenario Recognition \& Solution Mapping}

Customer service scenarios are determined by our two-stage scenario recognition model, which are represented as the "parameters" for determining solutions. It matches the customer utterance with the best service scenario by solving a pairwise text-match task. Taking the case of "complaining the bought shoes" as an example, the recognized best scenario is "returning the commodities (shoes)". 

Given a determined customer service scenario determined by our two-stage scenario recognition framework, the customer service solution can be determined and selected based on the scenario-solution mapping table formulated by the e-commerce company itself according to its business strategies. The customer service solution can be a customized one-to-one mapping from the customer service scenario. Specifically, the solution can be a standard service manual, a road-map, or just a predefined answer. For instance, the above-determined scenario is mapped to the solution of how to apply for the refund of the shoes.   

\section{Experimental Evaluation}
\label{sec:experiment}
Our experiment was conducted in the real customer service of Alibaba Group. We implemented ICS-Assist as an enterprise-level service system. Over 12,000 customer service staff use ICS-Assist to serve for over 230,000 cases per day on average, and this procedure lasted for over $6$ months. In our experimental environment, the queries made by end customers are dispensed to dedicated query processing servers by the query router. Each query processing server encapsulates and passes the query to the service scenario recognition model in our ICS-Assist to predict the service scenarios. The recognized scenarios are then mapped with the service solutions, which are sent to customers by customer service staff. 


\subsection{Experimental Procedure}
The experimental procedure consists of three parts:
\begin{enumerate}
    \item We apply our proposed ``Panel-Student" model and the baseline models \cite{rao2019bridging,chen2017enhanced,zhou-etal-2018-multi,DBLP:conf/iclr/GongLZ18} on the public dataset Quora \cite{quora}, and compare the performance among them. For the baseline models, we reproduce them according to their best hyper-parameter settings. For our proposed model, we set the convolution kernel widths from $2$ to $5$, and the output channel numbers are all $64$. The layer number of the MLP module in equation \ref{eq:mlp} is $3$. The dropout rate is $0.2$ and the L2 regularization coefficient is $0.05$. The activation function for all the layers is ReLu. The optimizer is Adam \cite{kingma2014adam} with a constant learning rate of 1e-4, decay rate $\beta_1$ of $0.9$, and $\beta_2$ of $0.999$.
    \item We also conduct similar experiments on our historical dataset. The only difference is that, for our proposed model, we use an additional neural model to handle the handcrafted multi-aspect features to construct the final hybrid model, we adopt a similar architecture like a Wide-Deep model \cite{cheng2016wide}, and combine it with the text-based features from the TextCNN model. The training for the final model follows a two-stage paradigm: 1) We freeze the parameters within TextCNN model and train the DNN with a constant learning rate of 1e-3 until convergence; 2) We make TextCNN's parameters trainable and restart the training phase with an exponential decay learning rate (initial learning rate: 1e-4; decay rate: $0.95$; decay step: $10000$).

\item Since the purpose of the two steps above is to demonstrate the superiority of our proposed ``Panel-Student" model, we replace this hybrid model in ICS-Assist with each of the state-of-the-art baseline models \cite{rao2019bridging,chen2017enhanced,zhou-etal-2018-multi,DBLP:conf/iclr/GongLZ18} to create several variants of ICS-Assist and compare our proposed ICS-Assist (with the ``Panel-Student" model)  with these ICS-Assist variants as well as the manual method against the $5$ business evaluation metrics (SAR, SCR, AST, CSR, and BCR) mentioned in Section \ref{subsec:business_req_for_cs}. 
\end{enumerate}

\subsection{Experimental Results}
Table \ref{tab:tech_exp_result_quora} shows the comparison between the performance (accuracy, precision, recall, f1-score, and latency) of our proposed Panel-Student model (PS model henceforth) and the other existing models on the Quora dataset. Due to the less execution complexity of our proposed model compared to other models, the accuracy, precision, recall and f1-score of our PS model are slightly less than hcan-hybrid, hcan-only rm \cite{rao2019bridging}, dam \cite{zhou-etal-2018-multi}, diin \cite{DBLP:conf/iclr/GongLZ18} and esim-seq \cite{chen2017enhanced}, and slightly better than hcan-only sm \cite{rao2019bridging}, but our PS model’s latency is much lower than these models. We also implement the panel-student model with a single teacher (i.e. BERT, XLNet, and ELMO), and obtain three Teacher-Student models (TS models henceforth). The performance of these three TS models is also worse than our PS model. 
As such, our model is the best one among all the models.

\begin{table}[h]
\caption{Model performance comparison on Quora dataset}
\label{tab:tech_exp_result_quora}
\vspace{-2mm}
\centering
\begin{tabular}{l|lllcr}
\toprule
model&accuracy&precision&recall&f1&latency(ms)\\
\midrule
hcan - hybrid&0.831&0.832&0.830&0.831&81\\
hcan - only sm&0.791&0.791&0.791&0.791&73\\
hcan - only rm&0.821&0.824&0.817&0.820&21\\
dam&0.855&0.856&0.854&0.855&109\\
diin&0.873&0.877&0.867&0.872&151\\
esim - seq&0.843&0.846&0.839&0.842&95\\
\midrule
TS - BERT&0.791&0.783&0.792&0.787&15\\
TS - XLNet&0.807&0.795&0.809&0.802&15\\
TS - ELMO&0.781&0.769&0.759&0.764&13\\
our PS model&0.811&0.807&0.819&0.813&11\\
\bottomrule
\end{tabular}
\vspace{-4mm}
\end{table}

Table \ref{tab:tech_exp_result_ours} shows the comparison between the performance of our proposed PS model and the existing state-of-the-art models using our dataset. Again, our PS model outperforms all the baseline models in terms of the overall model performance due to the less execution complexity of our model than others. 

\begin{table}[h]
\caption{Model performance comparison on our dataset}
\label{tab:tech_exp_result_ours}
\vspace{-2mm}
\centering
\begin{tabular}{l|lllcr}
\toprule
model&accuracy&precision&recall&f1&latency(ms)\\
\midrule
hcan - hybrid&0.878&0.876&0.879&0.877&198\\
hcan - only sm&0.845&0.847&0.841&0.844&186\\
hcan - only rm&0.850&0.848&0.852&0.850&53\\
dam&0.914&0.914&0.914&0.914&265\\
diin&0.894&0.899&0.887&0.893&387\\
esim - seq&0.930&0.932&0.928&0.930&241\\
\midrule
TS - BERT&0.871&0.874&0.877&0.875&28\\
TS - XLNet&0.878&0.884&0.889&0.886&26\\
TS - ELMO&0.853&0.851&0.861&0.856&25\\
our PS model&0.894&0.892&0.895&0.893&25\\
\bottomrule
\end{tabular}
\vspace{-4mm}
\end{table} 

Table \ref{tab:business_result}  shows the improvement rate of the business metrics of our proposed ICS-Assist (ICS-Assist (PS)) over the manual method and the variants of ICS-Assist with state-of-the-art models, including the teacher-student model using each single teacher model in our panel. Our ICS-Assist (PS) performs better than all the other variants of ICS-Assist by up to 16\%, 25\%, 6\%, 14\%, and 17\%, in terms of SAR (Solution Acceptance Rate), SCR (Solution Coverage Rate), AST (Average Service Time), CSR (Customer Satisfaction Rate) and BCR (Business Domain Catering Rate). Although our PS model's performance (e.g. f1-score) is slightly worse than other models (e.g. dam and esim-seq), our approach still performs better in the business metrics. This is because our PS model has lower latency than other models, and it enables customer service staff to timely utilize the recommended solutions. ICS-Assist (PS) increases SAR by 24\%, increases SCR by 34\%, decreases AST by 8\%, increases CSR by 19\%, and increases BCR by 22\%, compared to the manual method.

\begin{table}[h]
\caption{Business performance improvement results}
\label{tab:business_result}
\vspace{-3mm}
\centering
\begin{tabular}{l|lllll}
\toprule
ICS-Assist (PS) vs. &SAR&SCR&AST&CSR&BCR\\
\midrule
manual &24\%&34\%&8\%&19\%&22\%\\
ICS-Assist (dam)&13\%&19\%&4\%&5\%&7\%\\
ICS-Assist (hcan)&16\%&25\%&6\%&14\%&17\%\\
ICS-Assist (diin)&12\%&19\%&7\%&12\%&15\%\\
ICS-Assist (esim)&10\%&15\%&6\%&11\%&14\%\\
ICS-Assist (TS-BERT)&3\%&2\%&3\%&3\%&5\%\\
ICS-Assist (TS-XLNet)&1\%&2\%&2\%&2\%&7\%\\
ICS-Assist (TS-ELMO)&7\%&6\%&4\%&9\%&10\%\\
\bottomrule
\end{tabular}
\vspace{-3mm}
\end{table}
 
From the experimental results, we can conclude that our approach outperforms other automated state-of-the-art methods as well as the manual method in all the business metrics. The main reasons are as follows: 1) Our method assembles the three pre-trained language models (i.e. BERT, XLNet and ELMO) to distill a more generalizable model that creates better language representations for multiple business domains; 2) Our method takes multi-aspect features as the inputs for the scenario recognition model; 3) Our method employs a two-stage learning approach to maximize the validity of the recommended results, and it makes a reasonable prepossessing on the historical data to address its inevitable drawbacks related to quality, volume, and noise. 


\section{Threats to Validity}
\label{sec:threat}
The threats to validity are as follows: 1) The historical customer service data provided by Alibaba Group largely focus on the east Asian and southeast Asian countries, and the countries from other continents are relatively few. 2) The model training parameters with the PS model can be further tuned. The current parameters may not yield an optimized deep learning model because they may cause a local minimum instead of a global minimum. 3) The three representation learning-oriented models (BERT, XLNet, and ELMO) constitute the Panel, but more pre-trained language models could have been investigated. 

\section{Related Work}
\label{sec:relatedwork}
\subsection{Neural Text Matching Techniques}
One line of work related to our system is Neural Text Matching. Text Matching is a core task in many NLP and information retrieval applications, the mainstream of which can be divided into Semantic Matching (SM) and Relevance Matching (RM). Although both SM and RM are modelling similarities between two pieces of texts, SM emphasizes the semantic understanding and reasoning while RM focuses more on keyword matching signals. Typical SM tasks includes question answering \cite{chen2017reading}, paraphrase identification \cite{wang2017bilateral}, and natural language inference \cite{chen2017enhanced,DBLP:conf/iclr/GongLZ18}. RM models, such as DRMM \cite{guo2016deep}, Co-PACRR \cite{hui2018co} and MP-HCNN \cite{rao2019multi} are frequently used in IR applications like search engines to rank documents by relevance given a user query. In our work, both semantic and relevance matching technologies are involved in our model to enable more comprehensive language understanding and identify suitable service scenarios.

\subsection{Collaborative Filtering Techniques}
Because our system aims to recommend suitable service scenarios and solutions to the customer service staff, in this way the research work on recommendation Systems is also related to our work. Most recommendation systems are based on collaborative filtering, which learns a representation of user and item based on the rating matrix, and then predict the rating assigned a user given an unseen item. Currently, many recommendation systems adopt neural networks \cite{wang2015collaborative,he2017neural,ebesu2018collaborative} to learn a good dense representation and the interaction between the user and item and achieve the state-of-the-art performance. However in our scope,
mechanically matching the user and service scenario would ignore the customer's intention and requirement thus impair our service quality.

\subsection{Knowledge Distillation Techniques}
Researchers from the University of Waterloo try to transfer deep language representation like BERT to a lightweight neural network such as single-layer BiLSTM \cite{tang2019distilling}. But they do not employ multiple teachers’ knowledge to distil a simple student model. This experience motivates our multiple knowledge distilling.
In addition, The model Fitnets \cite{Romero2015FitNetsHF} is proposed by A. Romero. This model has extended the model compression idea and introduces the intermediate-level hints techniques to simplify a deeper and thinner student network with fewer parameters and better generalization. The new loss function is imported in hidden layers’ feature maps, which helps to reduce parameters in our work. Recently, the IBM researchers have proposed to train the student model from an ensemble of multiple teachers \cite{fukuda2017efficient}. They implemented different deep neural networks to train convolutional neural network acoustic models on a medium-sized speech corpus. The experimental results highlight that the proposed training techniques could increase a significant amount of knowledge to the student. Hence, our work also follows the idea of distilled learning by proposing a ``Panel-Student`` model.


\section{Conclusion \& Future Work}
\label{sec:conclusion}
Identifying proper customer service solutions is critical to e-commerce businesses. Existing service solution determination methods are usually unsatisfactory to end customers. This is because they are of low efficiency and unable to achieve runtime solution determination. Hence, this paper proposes an innovative framework, called ICS-Assist, to determine customer service solutions at runtime. We designed a novel two-stage learning model to identify customer service scenarios, which are mapped to end solutions. We implemented ICS-Assist and evaluated it in a 6-month real-world field study at Alibaba Group. The experimental results show that ICS-Assist improves the five business evaluation metrics (solution acceptance rate, solution coverage rate, average service time, customer satisfaction rate and business domain catering rate) by up to 16\%, 25\%, 6\%, 14\%, and 17\% respectively, compared to the state-of-the-art methods, and it outperforms the manual method by 24\%, 34\%, 8\%, 19\%, and 22\% respectively, in terms of the five business evaluation metrics. Our future work includes: 1) Explore more representation learning models for determining the members in the panel; 2) Design robust light-weight pre-trained models for customer services; 3) Investigate different customer service application areas such as finance.





%
%
%

\bibliographystyle{splncs04}
\bibliography{reference}

\end{document}